\documentclass[10pt]{article}
\usepackage{latexsym}
\usepackage{amssymb}
\usepackage{amsmath}
\usepackage{amscd}
\usepackage{amsthm}
\usepackage[left=2cm,top=2.5cm,right=2.5cm,bottom=1.5cm]{geometry}
\usepackage[dvips]{graphicx}
\usepackage{hyperref}
\begin{document}
\begin{center}
\large{\bf{Interacting Two-Fluid Viscous Dark Energy Models In Non-Flat Universe}} \\
\vspace{10mm}
\normalsize{Hassan Amirhashchi$^1$, Anirudh Pradhan$^2$, H. Zainuddin$^3$}\\
\vspace{5mm}
\normalsize{$^{1}$Young Researchers Club, Mahshahr Branch, Islamic Azad University, Mahshahr, Iran \\
{\it E-mail: h.amirhashchi@mahshahriau.ac.ir}} \\
\vspace{5mm}
\normalsize{$^{2}$Department of Mathematics, Hindu Post-graduate College, Zamania-232 331,
Ghazipur, India \\
{\it E-mail: pradhan@iucaa.ernet.in}} \\
\vspace{5mm}
\normalsize{$^{2,3}$Laboratory of Computational Sciences and Mathematical Physics, Institute for
Mathematical Research, University Putra Malaysia, 43400 UPM, Serdang, Selangor D.E., Malaysia \\
{\it E-mail: hisham@putra.upm.edu.my}}\\
\end{center}
\vspace{10mm}
\begin{abstract}
We study the evolution of the dark energy parameter within the scope of a spatially non-flat and isotropic
Friedmann-Robertson-Walker (FRW) model filled with barotropic fluid and bulk viscous stresses. We have
obtained cosmological solutions which exhibit without a big rip singularity. It is concluded that in both
non-interacting and interacting cases non-flat open universe crosses the phantom region. We find that during
the evolution of the universe, the equation of state (EoS) for dark energy $\omega_{D}$ changes from
$\omega^{eff}_{D} < -1$ to $\omega^{eff}_{D} > -1$, which is consistent with recent observations.
\end{abstract}
\section{INTRODUCTION}
Observations of distant Supernovae (SNe Ia) (Perlmutter  et al. 1997, 1998, 1999; Riess et al. 1998, 2000; Garnavich et al. 1998a,b; Schmidt et al. 1998; Tonry et al. 2003; Clocchiatti et al. 2006), fluctuation of cosmic microwave background radiation (CMBR) (de Bernardis et al. 1998; Hanany et al. 2000), large scale structure (LSS) (Spergel et al. 2003; Tegmark et al. 2004), sloan digital sky survey (SDSS)
(Seljak et al. 2005; Adelman-McCarthy et al. 2006), Wilkinson microwave anisotropy probe (WMAP) (Bennett. et al 2003) and Chandra x-ray observatory
(Allen et al. 2004) by means of ground and altitudinal experiments have established that our Universe is undergoing a late-time
accelerating expansion, and we live in a priviledged spatially flat Universe composed of approximately $4\%$ baryonic
matter, $22\%$ dark matter and $74\%$ dark energy. The simplest candidate for dark energy is the cosmological constant.
Recently, a great number of theme have been proposed to explain the current accelerating Universe, partly such as
scalar field model, exotic equation of state (EoS), modified gravity, and the inhomogeneous cosmology model. There are
several dark energy models which can be distinguished by, for instance, their EoS ($\omega = \frac{p_{de}}{\rho_{de}}$)
during the evolution of the universe. \\

The introduction of viscosity into cosmology has been investigated from different view points (Gr$\o$n 1990; Padmanabhan $\&$ Chitre 1987; Barrow 1986; Zimdahl 1996; Farzin et al. 2012).
Misner (1966, 1967;) noted that the ``measurement of the isotropy of the cosmic background radiation represents the
most accurate observational datum in cosmology''. An explanation of this isotropy was provided by showing that in
large class of homogeneous but anisotropic universe, the anisotropy dies away rapidly. It was found that the most
important mechanism in reducing the anisotropy is neutrino viscosity at temperatures just above $10^{10} K$ (when the
Universe was about 1 s old: cf. Zel'dovich and Novikov (Zel'dovich $\&$ Novikov 1971)). The astrophysical observations also indicate
some evidences that cosmic media is not a perfect fluid (Jaffe et al. 2005), and the viscosity effect could be concerned in
the evolution of the universe (Brevik $\&$ Gorbunova, 2005; Brevik et al. 2005; Cataldo et al. 2005). On the other hand, in the standard cosmological model, if
the EoS parameter $\omega$ is less than $-1$, so-called phantom, the universe shows the future finite time singularity
called the Big Rip (Caldwell et al. 2003; Nojiri et al. 2005) or Cosmic Doomsday. Several mechanisms are proposed to prevent the future big rip,
like by considering quantum effects terms in the action (Nojiri $\&$ Odintsov 2004; Elizalde et al. 2004), or by including viscosity effects for the Universe
evolution (Meng et al. 2007). A well known result of the FRW cosmological solutions, corresponding to universes filled with
perfect fluid and bulk viscous stresses, is the possibility of violating dominant energy condition (Barrow 1987, 1988; Folomeev $\&$ Gurovich 2008; Ren $\&$ Meng 2006; Brevikc $\&$ Gorbunovac 2005; Nojiri $\&$ Odintsov 2005).
Setare (Setare 2007a,b,c) and Setare and Saridakis (Setare $\&$ Saridakis 2000) have studied the interacting models of dark energy in
different context. Interacting new agegraphic viscous dark energy with varying $G$ has been studied by
Sheykhi and Setare (Sheykhi $\&$ Setare 2010). \\

Recently, Amirhashchi et al. (2011a,b); Pradhan et al. (2011); Saha et al. (2012) have studied
the two-fluid scenario for dark energy in FRW universe in different context. Very recently Singh and Chaubey (2012) have studied interacting dark energy in Bianchi type I space-time. Some experimental data implied that our
universe is not a perfectly flat universe and recent papers (Spergel et al. 2003; Bennett et al. 2003; Ichikawa et al. 2006) favoured a universe with spatial
curvature. Setare et al. (2009) have studied the tachyon cosmology in non-interacting and interacting cases in
non-flat FRW universe. Due to these considerations and motivations, in this Letter,  we study the evolution of
the dark energy parameter within the framework of a FRW open cosmological model filled with two fluids (i.e., barotropic
fluid and bulk viscous stresses). In doing so we consider both interacting and non-interacting cases.
\section{THE METRIC AND FIELD  EQUATIONS}
We consider the spherically symmetric Friedmann-Robertson-Walker (FRW) metric as
\begin{equation}
\label{eq1}
ds^{2} = -dt^{2} + a^{2}(t)\left[\frac{dr^{2}}{1 - kr^{2}} + r^{2}(d\theta^{2} +
\sin^{2}\theta d\phi^{2})\right],
\end{equation}
where $a(t)$ is the scale factor and the curvature constant $k$ is $-1, 0, +1$ respectively
for open, flat and close models of the universe.\\\\
The Einstein's field equations (with $8\pi G = 1$ and $c = 1$) read as
\begin{equation}
\label{eq2}
R^{j}_{i} - \frac{1}{2}R\delta^{j}_{i} = - T^{j}_{i},
\end{equation}
where the symbols have their usual meaning and $T^{j}_{i}$ is the two-fluid energy-momentum tensor due
to bulk viscous dark and barotropic fluids written in the form.
\begin{equation}
\label{eq3}
T^{j}_{i}=(\rho+\bar{p})u^{j}_{i}+\bar{p}g^{j}_{i},
\end{equation}
where
\begin{equation}
\label{eq4} \bar{p}=p-\xi u^{i}_{;i}
\end{equation}
and
\begin{equation}
\label{eq5} u^{i}u_{i} = -1,
\end{equation}
where $\rho$ is the energy density; $p$, the pressure; $\xi$, the bulk-viscous coefficient; and $u^{i}$,
the four-velocity vector of the distribution. Here after the semi-colon denotes covariant differentiation.\\\\
The expansion factor $\theta$ is defined by $\theta = u^{i}_{;i} = 3\frac{\dot{a}}{a}$. Hence Eq. (\ref{eq4})
leads to
\begin{equation}
\label{eq6} \bar{p} = p - 3\xi H,
\end{equation}
where $H$ is Hubble's constant defined by
\begin{equation}
\label{eq7} H = \frac{\dot{a}}{a}.
\end{equation}

Now with the aid of Equations (\ref{eq3})-(\ref{eq5}) and metric (\ref{eq1}), the surviving field equations
(\ref{eq2}) take the explicit forms
\begin{equation}
\label{eq8} \rho = 3\left(\frac{\dot{a}^{2}}{a^{2}}+\frac{k}{a^{2}}\right),
\end{equation}
and
\begin{equation}
\label{eq9} \bar{p} = -\left(\frac{\dot{a}^{2}}{a^{2}} + 2\frac{\ddot{a}}{a} + \frac{k}{a^{2}}\right).
\end{equation}
Also in space-time (\ref{eq1}) the Bianchi identity for the bulk-viscous fluid distribution $G^{;j}_{ij} = 0$
leads to $T^{;j}_{ij} = 0$ which yields
\begin{equation}
\label{eq10} \rho u^{i}+(\rho+\bar{p})u^{i}_{;i}
\end{equation}
which leads to
\begin{equation}
\label{eq11}\dot{\rho} + 3H(\rho + \bar{p})=0.
\end{equation}
Using Eq. (\ref{eq7}) in Eqs. (\ref{eq8}) and (\ref{eq9}) we get
\begin{equation}
\label{eq12} \rho=\left(\frac{3k}{A^{2}}e^{-2Ht}+3H^{2}\right),
\end{equation}
and
\begin{equation}
\label{eq13} \bar{p}=-\left(\frac{k}{A^{2}}e^{-2Ht}+3H^{2}\right),
\end{equation}
where $\bar{p} = p_{m} + \bar{p}_{D}$ and $\rho = \rho_{m} + \rho_{D}$. Here $p_{m}$ and $\rho_{m}$ are
pressure and energy density of barotropic fluid and $p_{D}$ and $\rho_{D}$ are pressure and energy
density of dark fluid respectively.\\

The equation of state (EoS) for the barotropic fluid $\omega_{m}$ and dark field $\omega_{D}$ are given by
\begin{equation}
\label{eq14}\omega_{m} = \frac{p_{m}}{\rho_{m}},
\end{equation}
and
\begin{equation}
\label{eq15}\omega_{D} = \frac{\bar{p}_{D}}{\rho_{D}},
\end{equation}
respectively.\\
From Eqs. (\ref{eq11})-(\ref{eq13}) we obtain
\begin{equation}
\label{eq16} \frac{\dot{\rho}}{3H} = \frac{2k}{a^{2}}e^{-2Ht}.
\end{equation}
Now we assume
\begin{equation}
\label{eq17} \rho = \alpha\theta^{2}~\mbox{or}~\rho=9\alpha H^{2},
\end{equation}
where $\alpha$ is an arbitrary constant. Eq. (\ref{eq17}) ensure us that our universe approaches homogeneity (Collins 1977). This condition has also been used by Banerjee et al. (1986) for deriving a viscous-fluid cosmological model with Bianchi type II space time.\\
Putting Eq. (\ref{eq17}) in Eq. (\ref{eq16}) and after integrating we get
\begin{equation}
\label{eq18} e^{-2Ht} = -\frac{3\alpha A^{2}}{2kt^{2}},
\end{equation}
which yields
\begin{equation}
\label{eq19} H = \frac{1}{2t}\ln\left(-\frac{2kt^{2}}{3\alpha A^{2}}\right),
\end{equation}
where $A$ is an arbitrary constant. From Eq. (\ref{eq19}), we observe that the condition given by (\ref{eq17}) restrict our study to the case when $k = -1 $
(i.e. only for open universe). In the following sections we deal with two cases, (i) non-interacting two-fluid model and
(ii) interacting two-fluid model.
\section{NON-INTERACTING TWO-FLUID MODEL}
In this section we assume that two-fluid do not interact with each other. Therefor, the general form of
conservation equation (\ref{eq11}) leads us to write the conservation equation for the dark and barotropic
fluid separately as,
\begin{equation}
\label{eq20}\dot{\rho}_{m} + 3\frac{\dot{a}}{a}\left(\rho_{m} + p_{m}\right) = 0,
\end{equation}
and
\begin{equation}
\label{eq21}\dot{\rho}_{D} + 3\frac{\dot{a}}{a}\left(\rho_{D} + \bar{p}_{D}\right) = 0.
\end{equation}
Integration Eq. (\ref{eq20}) and using (\ref{eq7}) leads to
\begin{equation}
\label{eq22}\rho_{m} = \rho_{0}a^{-3(1 + \omega_{m})}~\mbox{or}~\rho_{m}=\rho_{0}Be^{-3H(1 + \omega_{m})t},
\end{equation}
where $\rho_{0}$ is an integrating constant and $B=A^{-3(1 + \omega_{m})}$. By using Eq. (\ref{eq22}) in
Eqs. (\ref{eq12}) and (\ref{eq13}), we first obtain the $\rho_{D}$ and $p_{D}$ in term of Hubble's constant $H$ as
\begin{equation}
\label{eq23}\rho_{D} =\left(\frac{3k}{A^{2}}e^{-2Ht}+3H^{2}\right)-\rho_{0}Be^{-3H(1 + \omega_{m})t},
\end{equation}
and
\begin{equation}
\label{eq24} \bar{p}_{D} =\left(\frac{k}{A^{2}}e^{-2Ht}+3H^{2}\right)-\omega_{m}\rho_{0}B
e^{-3H(1 + \omega_{m})t}.
\end{equation}
\begin{figure}[htbp]
\centering
\includegraphics[width=8cm,height=8cm,angle=0]{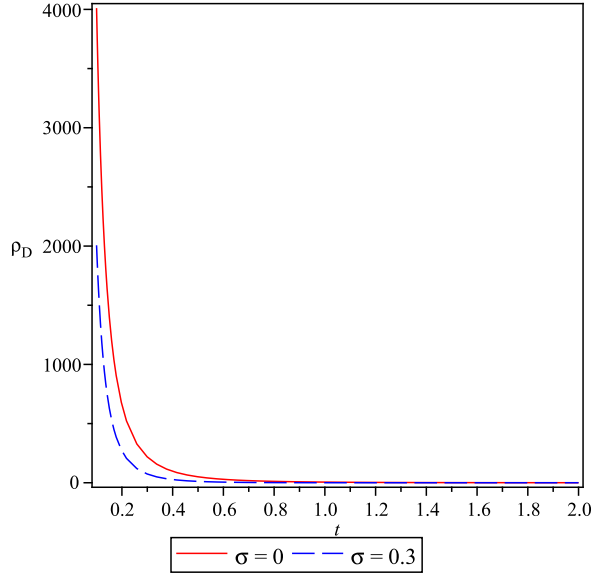}
\caption{The plot of $\rho_{D}$ vs $t$ for $\alpha = 0.1, A = 100, \omega_{m} = 0.5 $ in both non-interacting
and interacting two-fluid model}
\end{figure}
\begin{figure}[htbp]
\centering
\includegraphics[width=8cm,height=8cm,angle=0]{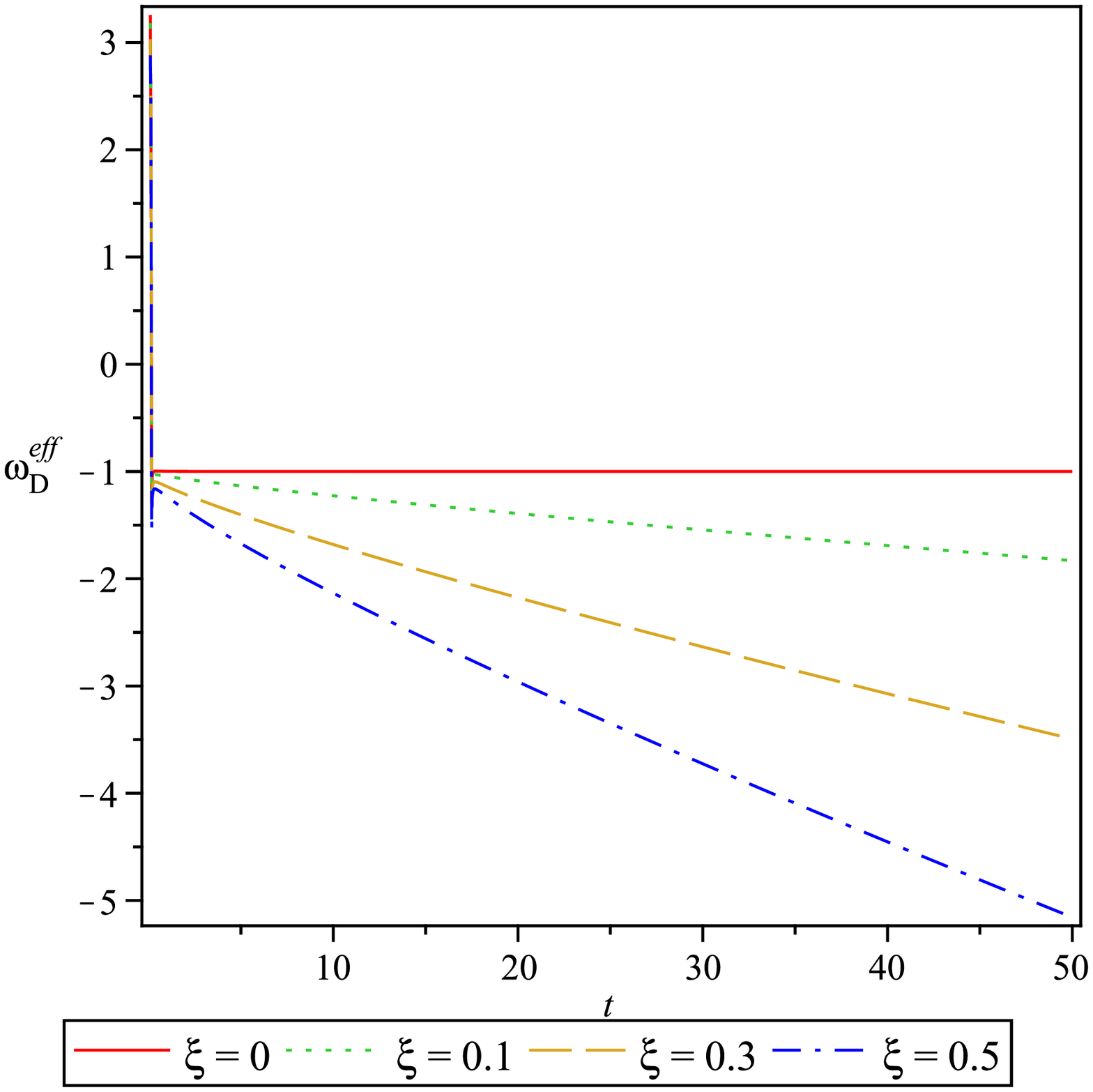}
\caption{The plot of EoS parameter $\omega^{eff}_{D}$ vs $t$ for $\rho_{0} = 10, \omega_{m} = 0.5,
\alpha = 0.01, B=1$ in non-interacting two-fluid model}
\end{figure}
respectively. By using Eqs. (\ref{eq23}) and (\ref{eq24}) in Eq. (\ref{eq15}), we can find the EoS of
dark energy in term of time as
\begin{equation}
\label{eq25}\omega_{D} = -\frac{\left(\frac{k}{A^{2}}e^{-2Ht}+3H^{2}\right)+
\omega_{m}\rho_{0}Be^{-3H(1 + \omega_{m})t}}{\left(\frac{3k}{A^{2}}e^{-2Ht} + 3H^{2}\right) -
\rho_{0}Be^{-3H(1 + \omega_{m})t}} .
\end{equation}
Therefore the effective EoS parameter for viscous DE can be written as
\begin{equation}
\label{eq26}\omega^{eff}_{D} =\omega_{D}-\frac{3\xi H}{\rho_{D}}= -\frac{\left(\frac{k}{A^{2}}e^{-2Ht}+3H^{2}\right)+3\xi H +
\omega_{m}\rho_{0}Be^{-3H(1 + \omega_{m})t}}{\left(\frac{3k}{A^{2}}e^{-2Ht} + 3H^{2}\right) -
\rho_{0}Be^{-3H(1 + \omega_{m})t}} .
\end{equation}
The expressions for the matter-energy density $\Omega_{m}$ and dark-energy density $\Omega_{D}$ are given by
\begin{equation}
\label{eq27}\Omega_{m} = \frac{\rho_{m}}{3H^{2}} = \frac{4t^{2}\rho_{0}Be^{-\frac{3}{2}\ln(\frac{2t^{2}}
{3\alpha A^{2}})(1 + \omega_{m})}}{3\ln^{2}(\frac{2t^{2}}{3\alpha A^{2}})},
\end{equation}
and
\begin{equation}
\label{eq28}\Omega_{D} =  \frac{\rho_{D}}{3H^{2}} = -\frac{6\alpha}{\ln^{2}(\frac{2t^{2}}{3\alpha A^{2}})}
+ 1-\frac{4t^{2}\rho_{0}Be^{-\frac{3}{2}\ln(\frac{2t^{2}}{3\alpha A^{2}})(1 + \omega_{m})}}
{3\ln^{2}(\frac{2t^{2}}{3\alpha A^{2}})},
\end{equation}
\begin{figure}[htbp]
\centering
\includegraphics[width=8cm,height=8cm,angle=0]{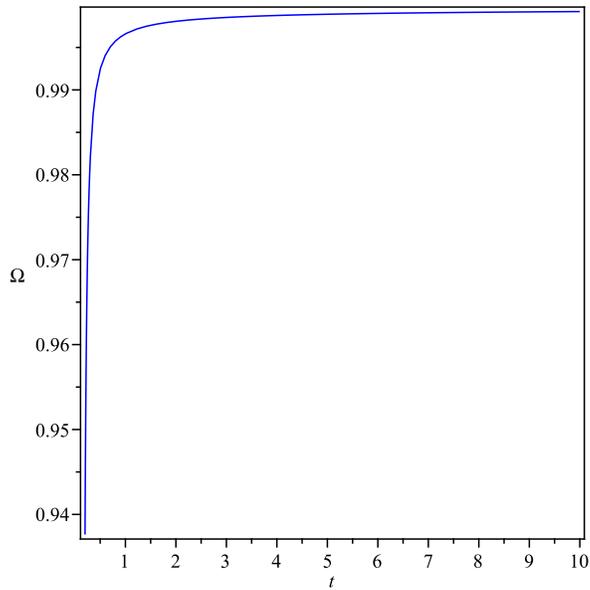}
\caption{The plot of density parameter ($\Omega$) vs $t$ for $A=1, \alpha = 0.01$ in non-interacting
two-fluid model}
\end{figure}
respectively. Adding Eqs. (\ref{eq27}) and (\ref{eq28}), we obtain
\begin{equation}
\label{eq29}\Omega = \Omega_{m} + \Omega_{D} = -\frac{6\alpha}{\ln^{2}(\frac{2t^{2}}{3\alpha A^{2}})} + 1.
\end{equation}
From the right hand side of Eq. (\ref{eq29}), it is clear that for open universe, $\Omega < 1$ but at late time we
see that  $\Omega \to 1$ i.e. the flat universe scenario. This result is also compatible with the observational
results. Since our model predicts a flat universe for large times and the present-day universe is very close to flat,
so being flat, the derived model is thus compatible with the observational results. \\\\
Fig. $1$ depicts the energy density of DE ($\rho_{D}$) versus $t$. From this figure, we observe that ($\rho_{D}$),
in both non-interacting and interacting cases, is a decreasing function of time and approaches a small positive
value at late time and never go to infinity. Thus, in both cases the universe is free from big rip. \\\\
The behavior of EoS for DE in term of cosmic time $t$ is shown in Fig. $2$. It is observed that for
open universe, the $\omega^{eff}_{D}$ is an decreasing function of time, the rapidity of its decrease at the early
stage depends on the larger value of bulk viscous coefficient. The EoS parameter of the DE begins in non-dark
($\omega_{D} > -\frac{1}{3}$) region at early stage and cross the phantom divide or cosmological constant
($\omega_{D} = -1$) region and then pass over into phantom ($\omega_{D} < -1$) region. The property of DE is a
violation of the null energy condition (NEC) since the DE crosses the Phantom Divide Line (PDL), in particular
depending on the direction (Rodrigues 2008; Kumar $\&$ Yadav 2011; Pradhan $\&$ Amirhashchi 2011). In theory, despite the observational constraints, extensions of
general relativity are the prime candidate class of theories consistent with PDL crossing (Nesseris $\&$ Perivolaropoulos 2007). On
the other hand, while the current cosmological data from SN Ia (Supernova Legacy Survey, Gold Sample of Hubble
Space Telescope) (Riess et al. 2004; Astier et al. 2006). CMB (WMAP, BOOMERANG) (Komatsu et al. 2009; MacTavish et al. 2006) and large scale structure
(SDSS) (Eisenstein et al. 2005) data rule out that $\omega_{D} \ll -1$, they mildly favour dynamically evolving DE
crossing the PDL (see Rodrigues 2008; Kumar $\&$ Yadav 2011; Pradhan $\&$ Amirhashchi 2011; Nesseris $\&$ Perivolaropoulos 2007; Zhao et al. 2007; Coperland et al. 2006) for theoretical and observational status of crossing
the PDL). Thus our DE model is in good agreement with well established theoretical result as well as the
recent observations. From Fig. $2$, it is observed that in absence of viscosity (i.e. for $\xi = 0$),
the universe does not cross the PDL but approaches to cosmological constant ($\omega_{D} = -1$) scenario.
Thus, it clearly indicates the impact of viscosity on the evolution of the universe. \\\\
The variation of density parameter ($\Omega$) with cosmic time $t$ for open universe
has been shown in Fig. $3$. From the figure, it can be seen that in an open universe, $\Omega$ is an increasing
function of time and at late time, it approaches to the flat universe's scenario.
\section{INTERACTING TWO-FLUID MODEL}
In this section we consider the interaction between dark viscous and barotropic fluids. For this purpose we can write
the continuity equations for barotropic and dark viscous fluids as
\begin{equation}
\label{eq30}\dot{\rho}_{m} + 3\frac{\dot{a}}{a}(\rho_{m} + p_{m}) = Q,
\end{equation}
and
\begin{equation}
\label{eq31}\dot{\rho}_{D} + 3\frac{\dot{a}}{a}(\rho_{D} + \bar{p}_{D}) = -Q,
\end{equation}
where the quantity $Q$ expresses the interaction between the dark components. Since we are interested in
 an energy transfer from the dark energy to dark matter, we consider $Q > 0$ which ensures that the second
law of thermodynamics is fulfilled (Pavon $\&$ Wang 2009). Here we emphasize that the continuity Eqs. (\ref{eq11}) and
(\ref{eq30}) imply that the interaction term ($Q$) should be  proportional to a quantity with units of
inverse of time i.e $Q \propto \frac{1}{t}$. Therefor, a first and natural candidate can be the Hubble
factor $H$ multiplied with the energy density. Following Amendola et al. (2007) and Gou et al.
(2007), we consider
\begin{equation}
\label{eq32}Q = 3H \sigma \rho_{m},
\end{equation}
where $\sigma$ is a coupling constant. Using Eq. (\ref{eq32}) in Eq. (\ref{eq30}) and after integrating,
we obtain
\begin{equation}
\label{eq33}\rho_{m} = \rho_{0}a^{-3(1 + \omega_{m} - \sigma)} ~ \mbox{or} ~ \rho_{m} =
\rho_{0}Be^{-3H(1 + \omega_{m} - \sigma)t}.
\end{equation}
By using Eq. (\ref{eq33}) in Eqs. (\ref{eq12}) and (\ref{eq13}), we again obtain the $\rho_{D}$ and $p_{D}$
in term of Hubble's constant $H$ as
\begin{equation}
\label{eq34}\rho_{D} = \left(\frac{3k}{A^{2}}e^{-2Ht} + 3H^{2}\right) - \rho_{0}Be^{-3H(1 + \omega_{m} -
\sigma)t},
\end{equation}
and
\begin{equation}
\label{eq35} \bar{p}_{D} = \left(\frac{k}{A^{2}}e^{-2Ht} + 3H^{2}\right)- (\omega_{m} -
\sigma)\rho_{0}Be^{-3H(1 + \omega_{m} - \sigma)t},
\end{equation}
respectively. By using Eqs. (\ref{eq34}) and (\ref{eq35}) in Eq. (\ref{eq15}), we can find the EoS of
dark energy in term of time as
\begin{equation}
\label{eq36}\omega_{D} = -\frac{\left(\frac{k}{A^{2}}e^{-2Ht} + 3H^{2}\right)+
(\omega_{m} - \sigma)\rho_{0}Be^{-3H(1 + \omega_{m} - \sigma)t}}{\left(\frac{3k}{A^{2}}e^{-2Ht} +
3H^{2}\right) - \rho_{0}Be^{-3H(1 + \omega_{m} - \sigma)t}}.
\end{equation}
Again we can write the effective EoS parameter of viscous DE as
\begin{equation}
\label{eq37}\omega^{eff}_{D} = -\frac{\left(\frac{k}{A^{2}}e^{-2Ht} + 3H^{2}\right) - 3\xi H +
(\omega_{m} - \sigma)\rho_{0}Be^{-3H(1 + \omega_{m} - \sigma)t}}{\left(\frac{3k}{A^{2}}e^{-2Ht} +
3H^{2}\right) - \rho_{0}Be^{-3H(1 + \omega_{m} - \sigma)t}}.
\end{equation}
\begin{figure}[htbp]
\centering
\includegraphics[width=8cm,height=8cm,angle=0]{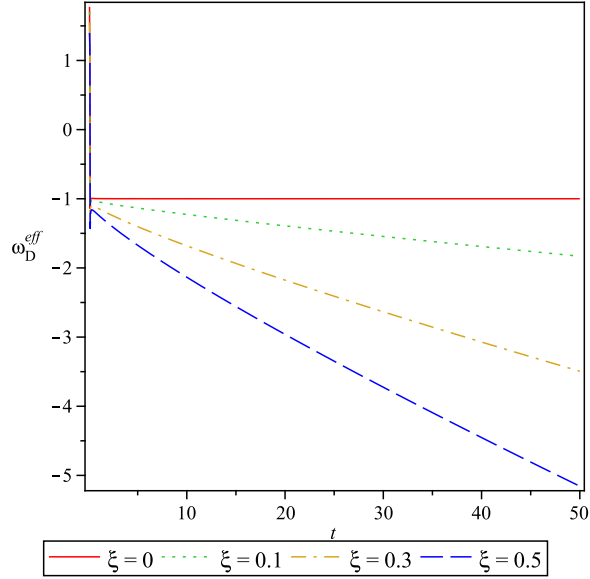}
\caption{The plot of EoS parameter $\omega^{eff}_{D}$ vs $t$ for $\rho_{0} = 10, \omega_{m} = 0.5, \alpha = 0.01,
B = 1, \sigma = 0.3 $ in interacting two-fluid model}
\end{figure}

The expressions for the matter-energy density $\Omega_{m}$ and dark-energy density $\Omega_{D}$ are given by
\begin{equation}
\label{eq38}\Omega_{m} = \frac{\rho_{m}}{3H^{2}} = \frac{4t^{2}\rho_{0}Be^{-\frac{3}{2}\ln(\frac{2t^{2}}
{3\alpha A^{2}})(1 + \omega_{m} - \sigma)}}{3\ln^{2}(\frac{2t^{2}}{3\alpha A^{2}})},
\end{equation}
and
\begin{equation}
\label{eq39}\Omega_{D} =  \frac{\rho_{D}}{3H^{2}} = -\frac{6\alpha}{\ln^{2}(\frac{2t^{2}}{3\alpha A^{2}})}
+ 1 - \frac{4t^{2}\rho_{0}Be^{-\frac{3}{2}\ln(\frac{2t^{2}}{3\alpha A^{2}})(1 + \omega_{m} - \sigma)}}
{3\ln^{2}(\frac{2t^{2}}{3\alpha A^{2}})},
\end{equation}
respectively. Adding Eqs. (\ref{eq38}) and (\ref{eq39}), we obtain
\begin{equation}
\label{eq40}\Omega = \Omega_{m} + \Omega_{D} = -\frac{6\alpha}{\ln^{2}(\frac{2t^{2}}{3\alpha A^{2}})} + 1,
\end{equation}
which is the same expression as in previous case of non-interacting two-fluid. Fig. $4$ shows a plot of EoS parameter
($\omega^{eff}_{D}$) versus $t$. The characteristic of $\omega^{eff}_{D}$ in this case is the same as in the previous case.
\section{CONCLUSION}
In this Letter, we have studied the evolution of dark energy parameter within the frame work of an open FRW space-time
filled with barotropic and bulk viscous dark fluid. In both non-interacting and interacting cases, we have observed
that for all values of bulk viscous coefficient, the universe has transition from non-dark region
($\omega^{eff}_{D} > -\frac{1}{3}$) to phantom region ($\omega^{eff}_{D} < -1$). In summary, we have investigated the possibility of
constructing a two-fluid dark energy models which have the equation of state ($\omega^{eff}_{D}$) crossing - 1 by using the
two-fluid (barotropic and bulk viscous dark fluid) naturally. Therefore, the two-fluid scenario discussed in the present
paper is a viable candidate for dark energy. It is also worth mentioned here that in both interacting and non-interacting
cases, our models are free from big rip. \\
\section*{ACKNOWLEDGMENT}
This work has been supported by the FRGS Grant by the Ministry of Higher Education, Malaysia under the
Project Number 02-10-10-969 FR. H. Amirhashchi \& A. Pradhan also thank the Laboratory of Computational
Sciences and Mathematical Physics, Institute for Mathematical Research, Universiti Putra Malaysia for providing
facility where this work was done.


\begin{thebibliography}{000}
\bibitem {ref1}
Adelman-McCarthy, J. K., et al. 2006, Astrophysical Journal Supplement, 162, 38
\bibitem {ref2}
Allen, S. W., et al. 2004, Monthly Notices of the Royal Astronomical Society,353, 457
\bibitem {ref3}
Amendola, L., et al. 2007, Phys. Rev. D, 75, 083506
\bibitem {ref4}
Amirhashchi, H., Pradhan, A $\&$ Saha, B. 2011a, Chinese Physics Letters,28, 039801
\bibitem {ref5}
Amirhashchi, H., Pradhan, A., $\&$ Zainuddin, H. 2011b, International Journal of Theoretical Physics,50, 3529
\bibitem {ref6}
Astier, P., et al. 2006, Astronomy $\&$ Astrophysics, 447, 31
\bibitem {ref7}
Banerjee, A., Duttachoudhury, S. B., $\&$ Sanyal, A. K. 1986, General relativity $\&$ Gravitation, 18, 461
\bibitem {ref8}
Barrow, J. D. 1986, Physics Letters B, 180, 335
\bibitem {ref9}
Barrow, J. D. 1987, Physics Letters B, 180, 335
\bibitem {ref10}
Barrow, J. D. 1988, Nuclear Physics B, 310, 743
\bibitem {ref11}
Bennett, C. L., et al. 2003, Astrophysical Journal Supplement, 148,  1
\bibitem {ref12}
Brevikc, I., $\&$ Gorbunovac, O. 2005, General relativity $\&$ Gravitation, 37, 2039
\bibitem {ref13}
Brevik, I., Gorbunova. O., $\&$ Shaido, Y. A. 2005, International Journal of Theoretical Physics D, 14, 1899
\bibitem {ref14}
Caldwell, R. R., Kamiionkowski, M., $\&$ Weinberg, N. N. 2003, Physical Review Letters, 91, 07301
\bibitem {ref15}
Cataldo, M., Cruz, N., $\&$ Lepe, S. 2005, Physics Letters B, 619, 5
\bibitem {ref16}
Clocchiatti, A., et al. 2006, Astrophysical Journal, 642, 1
\bibitem {ref17}
Collins, C. B. 1977, Journal of Mathematical Physics, 18, 2116
\bibitem {ref18}
Coperland, E. J., et al. 2006, International Journal of Modern Physics D, 15, 1753
\bibitem {ref19}
De Bernardis, P., et al. 1998, Nature, 391, 5
\bibitem {ref20}
Eisenstein, E., et al. 2005, Astrophysical Journal, 633, 560
\bibitem {ref21}
Elizalde, E., Nojiri, S., $\&$ Odintsov, S. D. 2004, Phys. Rev. D, 70, 0343539
\bibitem {ref22}
Farzin, A., et al. 2012, RAA, 12, 26
\bibitem {ref23}
Folomeev, V., $\&$ Gurovich, V. 2008, Physics Letters B, 661, 75
\bibitem {ref24}
Garnavich, P., et al. 1998, Astrophysical Journal, 493, L53
\bibitem {ref25}
Garnavich, P. M., et al. 1998, Astrophysical Journal, 509, 74
\bibitem {ref26}
Gr$\o$n, $\O$. 1990, Astrophysics $\&$ Space Science, 173, 191
\bibitem {ref27}
Guo, Z. K., Ohta, N., $\&$ Tsujikawa, S. 2007, Phys. Rev. D, 76, 023508
\bibitem {ref28}
Hanany, S., et al. 2000, Astrophysical Journal, 493, L53
\bibitem {ref29}
Ichikawa, K., et al. 2006, Journal of Cosmology and Astroparticle Physics,0612, 005
\bibitem {ref30}
Jaffe, T. R., et al. 2005, Astrophysical Journal, 629, L1-L4
\bibitem {ref31}
Komatsu, E., et al. 2009, Astrophysical Journal Supplement Series, 180, 330
\bibitem {ref32}
Kumar, S., $\&$ Yadav, A. K. 2011, Modern Physics Letters A, 26, 647
\bibitem {ref33}
MacTavish, C. J., et al. 2006, Astrophysical Journal, 647, 799
\bibitem {ref34}
Meng, X. H., Ren, J., $\&$ Hu, M. 2007, Communications in Theoretical Physics, 27, 379
\bibitem {ref35}
Misner, C. W. 1966, Astrophysical Journal, 151, 431
\bibitem {ref36}
Misner, C. W, 1967, Physical Review Letters, 19, 533
\bibitem {ref37}
Nesseris, S., $\&$ Perivolaropoulos, L. 2007, Journal of Cosmology and Astroparticle Physics, 0701, 018
\bibitem {ref38}
Nojiri, S., $\&$ Odintsov, S. D. 2005, Phys. Rev. D, 72, 023003
\bibitem {ref39}
Nojiri, S., $\&$ Odintsov, S. D. 2004, Physics Letters B, 595, 1
\bibitem {ref40}
Nojiri, S., Odintsov, S. D., $\&$ Tsujikawa, S. 2005, Phys. Rev. D, 71, 063004
\bibitem {ref41}
Padmanabhan, T., Chitre, S. M. 1987, Phys. Lett. A, 120, 433
\bibitem {ref42}
Pavon, D. $\&$ Wang, B. 2009, General Relativity and Gravitation, 41, 1
\bibitem {ref43}
Perlmutter, S., et al. 1999, Astrophysical Journal, 517, 5
\bibitem {ref44}
Perlmutter, S., et al. 1998, Nature, 391 51
\bibitem {ref45}
Perlmutter, S., et al. 1997, Astrophysical Journal, 483, 565
\bibitem {ref46}
Pradhan, A., Amirhashchi, H. 2011, Modern Physics Letters A, 30, 2261
\bibitem {ref47}
Pradhan, A., Amirhashchi, H., $\&$ Saha, B. 2011, Astrophysics and Space Science, 333 343
\bibitem {ref48}
Ren, J., $\&$ Meng, X. H. 2006, Physics Letters B, 633, 1
\bibitem {ref49}
Riess, A. G., et al. 2004, Astrophysical Journal, 607, 665
\bibitem {ref50}
Riess, A. G., et al. 2000, Publications of the Astronomical Society of the Pacific, 112, 1284
\bibitem {ref51}
Riess, A. G., et al. 1998, Astronomical Journal, 116, 1009
\bibitem {ref52}
Rodrigues, D. C., 2008, Phys. Rev. D, 77, 023534
\bibitem {ref53}
Saha, B., Amirhashchi, H., $\&$ Pradhan, A. 2012, Astrophysics and Space Science, DOI:10.1007/s10509-012-1155-x
\bibitem {ref54}
Schmidt, B. P., et al. 1998. Astrophysical Journal, 507, 46
\bibitem {ref55}
Seljak, U., et al. 2005, Phys. Rev. D, 71, 103515
\bibitem {ref56}
Setare, M. R. 2007, Physics Letters B, 644, 99
\bibitem {ref57}
Setare, M .R. 2007, European Physical Journal C, 50, 991
\bibitem {ref58}
Setare, M. R. 2007, Physics Letters B, 654, 1
\bibitem {ref59}
Setare, M. R., $\&$ Saridakis, E. N. 2008, Physics Letters B, 668,
177
\bibitem {ref60}
Setare, M. R., Sadeghi, J., $\&$ Amani, R. R. 2009, Physics Letters B, 673, 241
\bibitem {ref61}
Sheykhi, A., $\&$ Setare, M. R. 2010, International Journal of theoretical Physics, 49, 2777
\bibitem {ref62}
Spergel, D. N.,  et . 2003, Astrophysical Journal Supplement Series, 148, 175
\bibitem {ref63}
Tegmark, M., et al. 2004, Phys. Rev. D, 69 103501
\bibitem {ref64}
Tonry, J. L., et al. 2003, Astrophysical Journal, 594 1
\bibitem {ref65}
Singh, T., $\&$ Chaubey, R. 2012, RAA,12, 473
\bibitem {ref66}
Zel'dovich, Ya. B., $\&$ Novikov, I. D. 1971, Relativistic
Astrophysics, Vol. 2, The University Chicago Press, Chicago, p.
519.
\bibitem {ref47}
Zhao, G. B., et al. 2007, Physics Letters B, 648, 8
\bibitem {ref68}
Zimdahl, W. 1996, Phys. Rev. D, 53, 53483
\end{thebibliography}
\end{document}